# Universal Payment Channels:
# An Interoperability Platform for Digital Currencies*

Mihai Christodorescu, Erin English, Wanyun Catherine Gu, David Kreissman, Ranjit Kumaresan,
Mohsen Minaei, Srinivasan Raghuraman, Cuy Sheffield, Arjuna Wijeyekoon, Mahdi Zamani

Visa, Inc.

*Abstract.* With the innovation of distributed ledger technology (DLT), often known as blockchain technology, there has been significant growth of digital tokens in the form of cryptocurrencies, stablecoins, and central bank digital currencies. As the number of DLT networks increases, each with varying design characteristics, the likelihood that transacting parties are on the same network decreases. Thus, it is crucial to facilitate payments that are universal across networks, scalable to massive loads, and highly available.

We envision a future payment network that may be built on top of DLT networks without being subject to their limitations on interoperability, scalability, and availability faced by DLT payment solutions today. Specifically, we propose a hub-and-spoke payment route, referred to here as Universal Payment Channels (UPC), that can be used to support digital token transfers of funds across different networks through payment channels. We further discuss the potential use cases of the UPC technology to support, and not complicate, an already robust digital payment ecosystem. Finally, through the paper, we share some future directions of the UPC technology.

## Table of Contents

---

# 1    Introduction

Consumers need to be confident that they can exchange money for goods and services through payments securely. Consumers also need to feel assured that the speed of authorization and settlement, as well as consumer protection are robust. In a digital world, all these attributes remain highly relevant. Digital transactions that take place in the financial system are recorded by some type of digital ledger. Such a digital ledger is used to track the balances of the system's users and serves as a digital bulletin board, where all transactions in the system are posted [5].

With the innovation of distributed ledger technology (DLT), also known as blockchain technology, we have seen significant growth of digital tokens. These tokens include cryptocurrencies such as bitcoin and ether, as well as stablecoins, which are private digital tokens with their value pegged one for one to some external underlying fiat currency or other asset. The transfer of digital tokens is processed over blockchain networks - the largest two blockchains being the Bitcoin network and the Ethereum network. However, there are some well-known limitations to the design of these major blockchain networks when assessing them for payment purposes. First, a payment rail requires high throughput to facilitate retail payment use cases. Some of the largest blockchain networks today face scalability challenges as they cannot process substantial quantities of transactions per second (TPS). For example, in April 2021 the Ethereum network's throughput was limited to 16.5 TPS [26]; on average, it takes five minutes for a single transaction to be confirmed and settled on the Ethereum network compared to 65,000 TPS on VisaNet [27]. These slow processing times lead to significant network congestion.

A second challenge that most blockchain networks face today is interoperability. Many digital tokens built on top of the Ethereum network are interoperable with one another because they are created by smart contracts that adhere to the same set of Ethereum token standards [22]. However, there are many more digital tokens built on different blockchains which are not interoperable by design, such as the case between Ethereum and Bitcoin networks. A lack of natural interoperability poses a challenge to the transfer of crypto assets from one network to another.

As the number of blockchain networks increases, each with unique design characteristics, the probability for parties of a transaction to be on the same network decreases. Thus, it is crucial to facilitate payments that are both off-chain (to save overhead and fees) and universal (to transact across networks). Envisioning a future payment network that may be built on top of DLT networks but without the limitations highlighted, we propose a payment route that can be used to support digital token transfers called *Universal Payment Channels* (UPC).

Payment channels [24],[25] are defined as a class of mechanisms for making blockchain payments faster, by diverting as many transactions as possible to an off-chain communication channel between the payment sender and the payment recipient. Once off-chain channel is created through a funding transaction that takes place on-chain, then any number of subsequent transactions can be performed off the blockchain, and finally when one or both parties agree, the channel is closed through an exit transaction on the blockchain. This design mitigates both the costs and the latency associated with on-chain operations, effectively amortizing the overhead of the funding and exit transactions (which must be on-chain) over many off-chain transactions. Several schemes for payment channels have been proposed and even deployed, unfortunately with different degrees of limitations when compared to on-chain transactions, pushing users to make a trade-off between performance and functionality.

While the Turing-completeness of smart contracts can make interoperability between payment channels relatively easy to achieve, it also makes such payment channels somewhat less universal as they could transact only with networks that support such smart-contract functionality. The UPC technology to be described in this paper is designed to be universally interoperable with different types of blockchains. This has the benefit of allowing the UPC to onboard and connect with additional new blockchains in a reasonably short span of time. Based on the observation that digital signature verification and hash-time locked contracts (HTLCs) represent a common feature in the cryptocurrency ecosystem, we ask the following question:

*Is it possible to design a secure, universal payment channel based solely on signature verification and hash time lock contracts?*

The challenge here is that HTLCs and digital signature verification are much less powerful than general-purpose smart contracts, in that they simply unlock a transaction for processing in the network if the right cryptographic proof (either a hash preimage or a valid signature) is provided (typically before a specified deadline).

The UPC technology is also designed to be scalable, and this is achieved by a hub-and-spoke model. Contrary to a point-to-point topology in which a sender will need to open a bilateral channel with every one of its respective receivers, under a hub-and-spoke model, a sender will only need to set up a single channel with the hub which then connects to every other spoke in the network to achieve scalability. However, UPC addresses this shortcoming by utilizing a hub-and-spoke model, making the network much more efficient than a point-to-point network. By contrast, if a transaction is generated by a party without an established channel with the receiving party in a point-to-point network, the transaction may need to be routed through several intermediary nodes that are already established to circuitously reach the receiving party. If any of the intermediary nodes is not available or does not have sufficient capacity to handle incoming requests at the time of the transaction process, this will delay the transaction. On the other hand, a hub-and-spoke model requires only the hub to be available and to always have sufficient capacity to process off-chain transactions, to ensure the overall payment network availability.

Looking forward to the rest of this whitepaper, Background provides background context to ledger technology and payment channels. Universal Payment Channels describes and formalizes the detailed operation of the UPC protocol, which relies only on digital signature verification and HTLC support from the underlying blockchain network, thus satisfying the universality requirements. Along the way, we show how to correct the previous payment-channel designs where the sender would provide both the funds and the proof to unlock and complete the transaction. In UPC, the sender initiates a transaction by providing the funds and the recipient completes a transaction by providing the cryptographic proof. This choice leads to a simpler and more natural protocol. Use Cases for UPC discusses use cases of UPC to support multi-chain, multi-currency transactions for both retail and wholesale purposes. We propose the UPC technology can be useful in the context of central bank digital currencies (CBDCs) to support cross-border payment flows between CBDCs that may run on different networks. We also propose that the UPC technology can play an important role between private stablecoins and public CBDCs by providing permissioned access for whitelisted stablecoins to be interoperable with CBDCs. Future Directions concludes by sharing future direction of our work.

## 2   Background

In this section, we provide the context and the underlying technology on which UPC is built.

### 2.1   Ledger Technology

A digital ledger may be implemented using a centralized database controlled by a trusted third party, a decentralized ledger with no central point of control, or something in between such as a permissioned ledger. Regardless of the degree of centralization, a digital ledger, which is a digital record and an assurance of a transfer of value, typically needs to be distributed (i.e., replicated) geographically. This inherent resiliency is unique to blockchain technology and can help mitigate crash failures and malicious corruptions of its nodes.

Under a DLT network, there consists of (1) a set of computers known as *nodes* that store the ledger data, (2) a communication network for the node(s) to receive transactions and possibly communicate with each other, and (3) a set of protocols that describe precisely how the nodes can process and store the transactions securely. A *consensus protocol* executed by the nodes guarantees that even large subsets of nodes cannot collude to maliciously modify the ledger. Furthermore, a stack of network protocols ensures reliable delivery of messages, and an identity mechanism prescribes how participants can obtain identities in the form of digital signature keys to join the consensus protocol and/or create transactions. Additionally, and perhaps most critically, the consensus protocol process improves the overall robustness and security of the network compared to a centralized network and mitigates the likelihood that a cyber threat actor can compromise, corrupt, or manipulate the network's integrity. In short, DLT can enhance information assurance — a key cybersecurity attribute.

Moreover, DLT can enable automatic verification of ledger events to external entities, and depending on the ledger design and governance procedures, can even do so to ensure a level of privacy. Relatedly, a DLT network may optionally be able to execute computer programs, called smart contracts. A smart contract execution may be initiated by either a transaction submitted to the ledger or another contract. In addition to transactions for transferring assets and for initiating smart contract executions, a ledger that supports smart contracts can further store program data, referred to as contract state, for future executions of the contract. Notably, developers could also architect smart contracts into the system to enforce governance or regulatory requirements.

Two major issues with DLT protocols are scalability and interoperability. To provide resiliency guarantees against system failures and malicious activities, DLT protocols usually create redundancy by replicating ledger data across multiple machines that are ideally distributed across different geographical locations. This unfortunately creates an inherent scalability challenge to ensure that all or most of the replicas are in sync via a consensus protocol, which unfortunately, imposes significant communication overhead on the network of machines. Moreover, consensus protocols are designed in a way to ensure certain guarantees in the system they operate in; these protocols are not recognized by nodes in other blockchain systems. This individuality of networks introduces an inherent challenge in communication between multiple systems. We note that DLT can be used as a mechanism to provide performance benefits to the underlying system, such as data localization and parallelization. There is a natural tension between decentralization and the performance gains of going distributed and we expect different DLTs to make distinct design choices to find the optimum tradeoff. Regardless of this tradeoff, we believe that scalability and interoperability challenges generally appear in some degree in any DLT system.

## 2.2 Payment Channels

Depending on the scalability requirements of cross-border payments, a cross-ledger protocol may authorize payments in two ways: *on-ledger* or *off-ledger*. An on-ledger payment entails writing the transaction directly on the ledger at the time of payment, while an off-ledger payment relies on a collateralized payment channel to authorize payments securely without writing on the DLT at the time of funds transfer. In practice, an on-ledger payment could take significantly longer to confirm due to the latency of consensus protocols and the potentially massive load on the blockchain networks. In contrast, off-ledger payments are confirmed instantly and can scale to a virtually unbounded load. In the next section, we propose payment routes referred to as *Universal Payment Channels* that amortizes the ledger overhead by making one-time deposits into a smart contract and then enabling payments recipients several times without writing on the ledger for each payment. We will discuss UPC in more detail in Universal Payment Channels.

### 2.2.1 Hash-Time Locked Contracts

To be interoperable with other ledgers, a ledger protocol may provide, at minimum, basic smart contract execution capabilities to support HTLCs. Fortunately, such contracts can be provided by both blockchain and non-blockchain-based ledgers. An HTLC provides the following functionalities:

1. Locking collateral funds on both ledgers to create a UPC channel, and;

2. Releasing the collateral funds as part of a final settlement of the channel, which could be initiated either automatically at the channel's expiry time or manually by any of the participants (e.g., in case of a dispute or manual channel termination).

More specifically, an HTLC provides two primitives: *timelocks* and *hashlocks*. A timelock is a primitive that allows a smart contract to restrict spending of some funds until a specified time in the future while a hashlock is a primitive that restricts spending of funds until a secret is revealed to the contract. Given a cryptographic hash function $H$, the secret is usually represented in the form of a hash preimage $x$, where $H(x)$ is provided to the contract as a commitment to $x$. The commitment allows the contract to ensure that the secret revealed by the committer later (e.g., settlement time) is mathematically bound to some promise the committer made to another party at an earlier time (e.g., transaction time). This is the core property of HTLCs that can be used to reduce counterparty risk in two-party transactions.

### 2.2.2 Other Payment Channel Solutions

We provide a brief comparison to previous payment channel proposals, namely Lightning, Raiden, and Polygon. Payment channel networks, also popularly known as layer-2, offer an integrated scaling solution to sending, routing, processing, and receiving off-chain payments through the network.

Lightning [16] is a payment channel network that supports Bitcoin. State updates after a successful payment corresponds to newly signed UTXO transactions that can be submitted directly to the Bitcoin blockchain in case of disputes. Lightning relies on HTLCs to enable multi-hop routing of payments across the Lightning network. It does not support Ethereum. Lightning can support up to 483 concurrent payments on a single channel [19].

Raiden [17] is a payment channel network that supports Ethereum and ERC-20 tokens. Raiden uses a single smart contract on the Ethereum blockchain to manage all the payment channels in its network. State updates after a successful payment corresponds to newly signed messages on the latest balance that can be submitted directly to the Raiden smart contract on the Ethereum blockchain. Raiden relies on HTLCs to enable multi-hop routing of payments across this network, while supporting 160 concurrent payments on a single channel [20]. Unlike UPC, Raiden does not espouse a hub-and-spoke model, and it uses a single on-chain contract to manage all bilateral payment channels.

Polygon [18] (previously Matic Network) is a protocol for building and connecting Ethereum-compatible blockchain networks. Polygon uses a proof-of-stake high throughput sidechain with a selected set of so-called block producers chosen for every checkpoint by a set of stakers. The proof of stake layer validates the blocks and publish periodic proofs of the blocks to the Ethereum mainnet. More recently, Polygon attempts to provide an integrated layer-2 solution for Ethereum.

## 3 Universal Payment Channels

We now describe UPC and show how it can provide a scalable, interoperable platform for digital currencies. UPC operates in a hub-and-spoke model, where clients register with a UPC hub to route their transactions to other clients. Note that this routing requires zero trust to be placed on the UPC hub (the UPC hub does not need to be trusted like a central intermediary).

Towards this end, the UPC protocol uses timelocks and hashlocks to minimize counterparty risks in a three-party model (payer-hub-payee), under a hub-and-spoke design. The immediate benefit of such a protocol is the ability to scale UPC to millions or even billions of users and/or transactions while imposing minimal liability on the UPC hub via a prefunded model and consequently reducing fees and complexities of cross-border payments. When used to route transactions between two different ledgers, the UPC protocol requires both ledgers to support the same hash function. This requirement is in place so that the corresponding smart contracts on the two ledgers can lock funds with the same hash value on both ledgers and unlock them with the secret tokens associated with the hash value. On the other hand, UPC does not demand that ledgers support the same digital signature scheme while any client-hub pair are required to agree on the same signature scheme so that they can authenticate each other's messages.

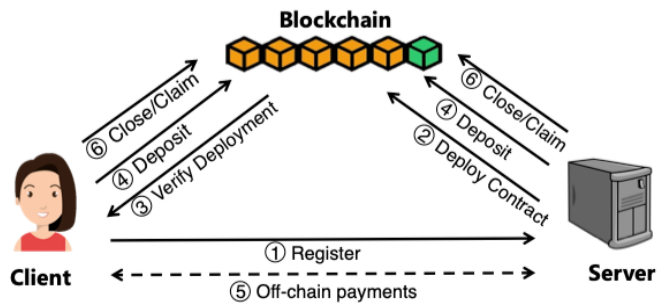

*Figure 1: The Main Stages of a UPC Channel*

We now outline the life cycles of a UPC client, a UPC channel, and a UPC transaction. A party who wishes to send funds through the UPC system will register with the server by providing their public key. A channel is *opened* between a registered client and the server when the server deploys an instance of the UPC smart contract on the ledger used by both parties and initializes the contract with channel parameters specific to the two parties. The channel is then pre-funded when the client and the server make *deposits* into the contract's address. We say the channel is *closed* when either or both parties decide to settle (i.e., finalize) their off-chain balances. These channel operations require on-chain transactions by one or both parties. Finally, a UPC payment transaction between two registered clients $A$ and $B$ is performed purely off-chain through $S$.

## 3.1   The UPC Hub

The UPC protocol facilitates payments through an entity, called the UPC *hub* (or server — we use the terms interchangeably), which acts as a gateway to receive payment requests from registered sending parties and routes them to registered recipient parties. The UPC hub is trusted to be highly available and process payment requests, and by design, its operation is fully transparent to any entity that can read the state of the two ledgers. Our protocol requires the UPC hub to authorize every payment that happens between the parties off the ledger. Thus, the UPC hub could check the validity of every payment. There may be multiple hubs running at the same time, connecting to the same ledger, and clients can register and transact with any of them.

## 3.2   Contract Deployment and Initial Deposit

In Figure 1, before a party can send or receive payments to/from the UPC hub, they need to register their identity (i.e., the public key obtained during the initial setup) with the UPC hub and open a pre-funded channel on the corresponding ledger jointly with the server, as shown in Step 1.

After the party is registered, the UPC contract is deployed on the ledger by the server (Step 2). The UPC contract consists of a common set of instructions to open, close (aka, settle), and dispute transactions, and must be written in the specific smart contract language supported by the ledger. This is a one-time process that is done by the UPC hub admin when on-boarding a new client. The smart contract guarantees the interests of all parties are incentive compatible. A dispute initiated by an honest party could result in forfeiture of some or all the deposits by the misbehaving party. The validity of the UPC contract execution and the security of the deposits held by the contract are all guaranteed by the underlying ledger.

After the client verifies the deployment of the contract by the server (Step 3), both parties transfer an agreed-upon amount of money, that they own on the ledger, into the UPC contract (Step 4). Next, we move on to the off-chain transactions made between the different clients of the UPC system (Step 5).

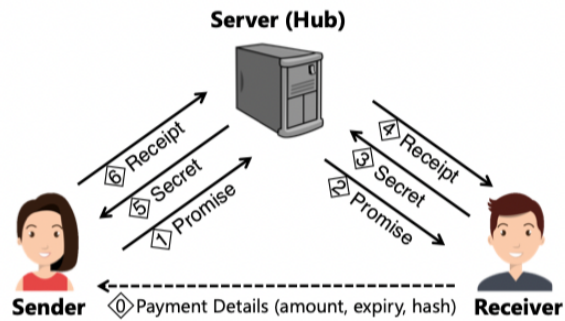

*Figure 2: The Steps in an UPC Off-chain Payment*

### 3.3   UPC Authorization & Pay

Once both parties register and open UPC channels with the UPC hub, they are ready to send/receive payments via off-ledger communication with the hub. A UPC transaction consists of three types of messages exchanged between the three parties:

1. **Promise:** A signed message promising a transaction amount conditioned on getting a secret from the counterparty before an expiration time. The message contains a hash of the secret that was previously presented to the sender by the recipient to "pull" the transaction.

2. **Secret:** A message containing a secret value generated by the recipient attesting that the payment promise was received and accepted.

3. **Receipt:** A message sent by a party after receiving a secret associated with their promise. The message is signed by the party and includes a credit value that denotes the aggregated amount of promises for which the corresponding secrets have been revealed.

Figure 2 shows how these messages are exchanged as part of a UPC three-party protocol. Every UPC transaction consists of two steps: authorization and payment. In the former step, the recipient sends to the sender (e.g., via a QR code shown on a point-of-sale device) a proposal akin to an invoice (Step 0). As part of the proposal, the recipient includes a hash of a secret that it generates uniquely at random for this transaction. The hash serves as a commitment by the recipient pledging that it will eventually (by a time set in the payment proposal, referred to as expiry) reveal the secret upon receiving a valid promise from the sender. Otherwise, the promise does not incur any liability for the sender (i.e., no money will be deducted from the sender's initial deposit in the settlement phase). After receiving the payment proposal, the sender creates a promise and sends it to the server (Step 1) who verifies the promise and creates a corresponding promise for the receiver (Step 2). Similarly, the receiver verifies the server's promise and proceeds to the Pay step.

The receiver begins the latter payment step by sending the secret to the server (Step 3) who verifies the secret and acknowledges it by sending back a receipt that includes the updated credit value (increased by the amount in the promise) of the receiver (Step 4). Next, the server forwards the secret to the sender (Step 5) who verifies and acknowledges the receipt similarly (Step 6). This ends the Auth & Pay step of the protocol.

### 3.4   UPC Settlement

All parties always maintain the latest signatures that they received off-ledger from the other parties during the Auth & Pay phase, so that they can go to the UPC contract on the ledger and submit their credit claims. Using the signatures submitted by the parties, the UPC contract can automatically calculate the final balance of each party considering their initial deposits and settle the channel. This closes the UPC channel, and the two parties must open a new channel if they wish to transact again. Note that the maximum amount of funds each party can spend in the UPC channel is equal to the amount that the party deposits into the contract added by the difference between the credit it receives and the credit it sends to the other party.

## 3.5    Event Handler

In addition to registered parties and the hub, UPC makes uses of an *event handler*, which runs in the background to observe the state of the channel and any unexpired promises. When the channel is active, it deletes expired promises and goes on-chain to claim promises *that are about to expire*. When the channel is closing, it goes on-chain to claim all promises and then proceeds to close the channel cooperatively. Finally, once the channel is cooperatively closed, the final settled amounts are transferred to the respective parties' accounts.

## 4    Use Cases for UPC

### 4.1    Cross-Border Payments for CBDCs

A CBDC represents a digital form of central bank liability issued by a central bank and intended as legal tender. While a CBDC system could bring efficiency to domestic economies partly through unified technologies for minting, distribution, and payment rails, envisioning similar unified models for cross-border payments (XBPs) between independent CBDC networks would be challenging.

Enabling cross border money movement of fiat today typically involves a network of corresponding banks. According to the International Monetary Fund (IMF), "A correspondent banking arrangement involves one bank (the correspondent) providing a deposit account or other liability accounts, and related services, to another bank (the respondent), often including its affiliates. The arrangement requires the exchange of messages to settle transactions by crediting and debiting those accounts" [14]. This model was adopted primarily because the bank used by the source and destination are the most familiar and specialized in their respective jurisdiction.

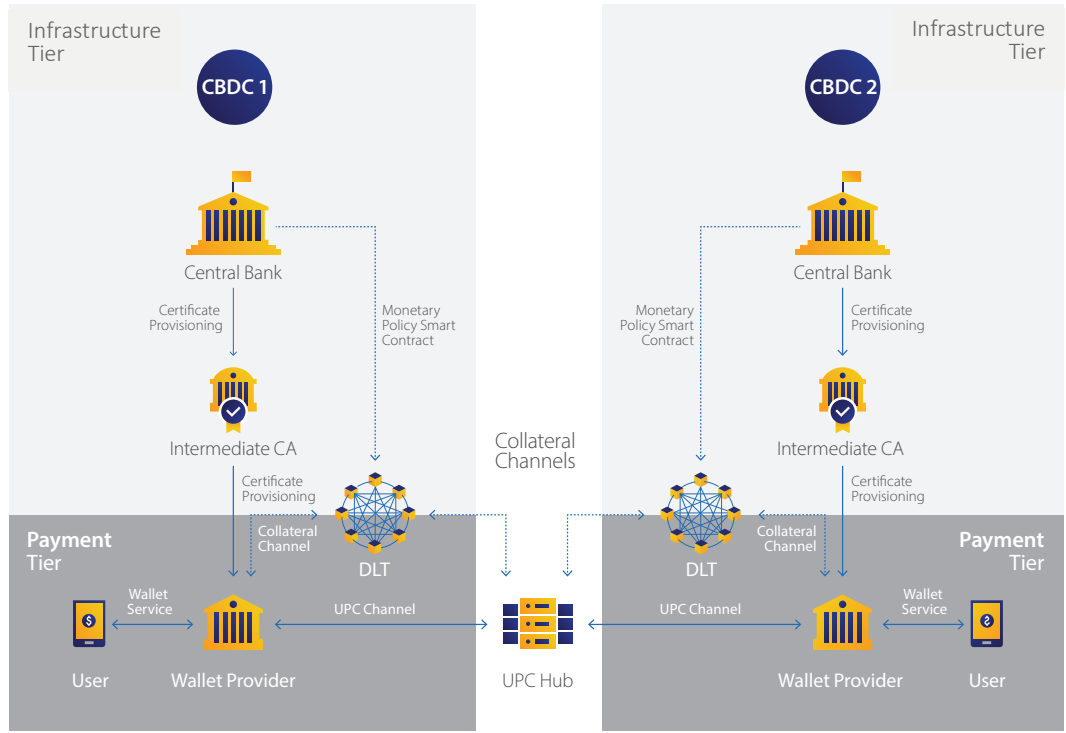

*Figure 3: The Two-Tier CBDC Model and Cross-Border CBDC Payments Using UPC*

In the case of a token-based CBDC, we frame a cross-border CBDC environment with a user with funds recorded on a CBDC ledger who wants to send a payment to another user on another CBDC ledger. We assume that the two CBDC ledgers are maintained by separate networks, use different ledger protocols, and reside in different jurisdictions. We further assume that both ledgers are implemented as DLTs, i.e., each ledger is replicated across multiple geographically distributed nodes for resiliency. We finally assume both DLTs support basic, standard smart contract execution that supports digital signature verification (e.g., ECDSA verification) and hash functions (e.g., SHA-3). Figure 3 shows the cross-border CBDC infrastructure using UPC in the context of the two-tier CBDC model described. As shown in the figure, each CBDC system allows the central bank to delegate the task of key provisioning to one or more intermediate certificate authorities (CAs) who provision keys on behalf of the central bank as *wallet providers*. In the context of a cross-border CBDC payments scenario like international remittances, where both the sender and the receiver rely on their respective banks (in this case, wallet providers) to send and receive funds on their behalf, UPC technology can be used to provide that bridge between two sets of CBDC systems. In addition, the sender may self-custody her funds by storing secret keys locally and authorizing her wallet provider via a digital signature to initiate an XBP. The wallet provider then submits the transaction to the UPC protocol. Depending on the frequency of XBP requests submitted to the wallet provider and the XBP service provided to the users, the wallet provider may submit a wholesale XBP through the UPC hub to tradeoff between the latency and the cost of XBPs.

Ultimately, given the complex nature of cross-border transactions today, processing for these transactions is more costly than processing for domestic transactions. Maintaining such an advanced, reliable, and robust network requires ongoing investment in technology, product development, risk mitigation, fraud detection capabilities, and regulatory compliance. It is critically important for a CBDC ecosystem to support, and not complicate, an already robust cross-border payment system [1] that markets have come to depend upon.

## 4.2    Marketplace for Digital Currencies

With the combination of scalability and interoperability features that the UPC technology provides, UPC hub can be a bridge connecting regulated stablecoins with CBDCs in the future. We envision that the development of this technology will significantly expand the utility of digital currencies as means of making digital payment across a network of businesses, consumers, and developers – whether C2B, B2B, or P2P.

On the wholesale level, as the previous section shows, moving money around the globe is often slow and expensive. UPC is designed to serve as a cross-border hub to transfer funds safely and efficiently in a programmable way. Reducing the settlement time in payments especially for large cross-border wholesale activities, UPC hub can provide a valuable functionality for businesses looking to upgrade supply chain efficiency or to improve treasury funds management.

On the retail level, UPC and its payment channels can support high throughput between different digital currencies enabling efficient transactions, in turn permitting for the adoption of digital currencies for everyday purchases and new use cases. UPC can streamline the payment experience in digital currencies for P2P transactions – digital currency payments can be confirmed in a matter of seconds. UPC also makes cross-border P2P transactions more cost effective, whether for a remittance or a purchase of a foreign good from your friend abroad. Moreover, UPC may enable new ways to pay. For instance, it is possible in the future to micro-tip your favorite artist or avatar in digital currencies in real-time across different platforms using a single digital wallet through UPC. UPC as an infrastructure has features that can be more fully developed to support new types of digital payments as they emerge, across digital payment platforms to transact between different digital assets and tokens.

## 5    Future Directions

### 5.1    Concurrent Transactions in UPC

The off-chain protocol described in Universal Payment Channels assumes that the transactions are submitted and processed in a serialized fashion, meaning that the client cannot initiate a new transaction unless the previous one has completed. This sequence is in place because when a payment receipt is sent, it includes a credit value that denotes the aggregate of the promise amounts for which the secret was received. As a result, when transactions are sent concurrently, a malicious party can deceive the contract by presenting a (promise, receipt) pair for which the amount of the promise is already included in the receipt. Such an attack can be mitigated by including an index/counter for the promises and receipts, such that old promises are invalidated by higher index values included in every receipt. Unfortunately, this forces off-chain transactions to happen sequentially, where no new promise will be accepted unless the previous one is satisfied (i.e., the corresponding receipt is received).

 To provide the maximum parallelization for a receiver that can process multiple promises at the same time, the UPC protocol would ideally allow the sender to submit multiple promises without waiting for each promise to be processed (we refer to this property as *non-blocking/concurrent* payments). It is critical to capture the list of pending promises in an efficient manner to attach them along with newly generated receipts. To address this issue, UPC could utilize cryptographic accumulators such as Merkle trees and/or RSA accumulators. This allows the reduction of the asymptotic bandwidth/fee overhead of inclusion proofs to a logarithm (e.g., for a Merkle tree) or a constant (e.g., for an RSA accumulator) in the number of pending promises. Finally, we note that similar ideas have previously been employed to handle concurrent payments in the design of the Raiden payment channel network.

## 5.2    Privacy

Most CBDCs will most likely require mechanisms to hide personal and/or financial information recorded on the DLT from unauthorized parties. This information includes both the amounts of transactions as well the identities of the sender and the receiver. We assume that the transaction information is encrypted in such a way that it can still be audited for validity and compliance by auditors while still preserving the privacy of the transaction. Adapting existing compliance mechanisms in non-CBDC flows to the CBDC world is still an open area that requires further research and development to improve efficiency, accuracy, and security. A summary of such efforts can be found in the Brooking's Institute's *Design Choices for CBDC* [5]. It is important to stress that most jurisdictions do not allow for anonymity in electronic payments to the same degree as in the cash world, which is likely to remain the case regardless of whether a CBDC involves an intermediary-based model or not. We appreciate the challenges navigating the privacy concerns of a digital currency, but anonymity also poses potential challenges such as the risk of invoking Gresham's Law of "bad money driving out good money."

As described in Universal Payment Channels the UPC protocol only records the final settlement amounts on the DLT which would be publicly visible. It thus automatically protects privacy of transactions at an individual level. The UPC Hub (which is an *authorized party*) does have visibility into the elements of individual payment transactions between the two end points. However, several prior works in the research literature have focused on privacy-preserving off-chain payments - building on and adapting such ideas to UPC is a direction for future research.

## 5.3    UPC-as-a-Service

As mentioned earlier, UPC relies on HTLCs to perform a trustless payment routing from $A$ to $B$ through the UPC Hub. Interestingly, the UPC payment from $A$ to $B$, takes the form of a standard HTLC payment (except it is now off-chain). Since on-chain HTLC payments are sufficient to enable interesting applications such as atomic swaps of cryptocurrencies [21], UPC, as is, can thus support a layer-2 implementation of any such application. Generalizing UPC conditional payments beyond HTLCs may further increase the range of applications for which UPC can act as a useful base layer service to support sophisticated layer-2 applications. We leave this for future research.

## 5.4    Liquidity Management

Due to the pre-funded model of UPC, it is important for the UPC hub and for UPC clients to manage liquidity on their UPC channels. Managing liquidity on layer-2 will require the ability to move liquidity without making on-ledger state changes between parties. Currently, UPC does not provide mechanisms for doing such money movement. We leave the design of such liquidity management mechanisms for future research.

## Acknowledgement

The authors would like to thank Karan Patel, Vinjith Nagaraja, Ming Xu, Benjamin Price, Akshay Kant, and Chad Harper for valuable comments and discussions.

## Disclaimer